\documentstyle[sprocl,epsfig]{article}
\bibliography{unsrt}
\def\be{\begin{equation}}
\def\beq{\begin{equation}}
\def\ee{\end{equation}}
\def\eeq{\end{equation}}
\def\bea{\begin{eqnarray}}
\def\eea{\end{eqnarray}}
\def\bcdot{{\bf\cdot}}

\font\tenbifull=cmmib10 
\font\tenbimed=cmmib10 scaled 800
\font\tenbismall=cmmib10 scaled 666
\def\boldsigma{\fam=9{\mathchar"711B } }

\def\dsl{\,\raise.15ex\hbox{/}\mkern-13.5mu D} 
\def\delsl{\raise.15ex\hbox{/}\kern-.57em\partial}
\def\Ksl{\hbox{/\kern-.6000em\rm K}}
\def\Asl{\hbox{/\kern-.6500em A}}
\def\Dsl{\hbox{/\kern-.6000em\rm D}} 
\def\Vsl{\hbox{/\kern-.6000em V}}
\def\gradsl{\hbox{/\kern-.6500em$\nabla$}}

\def\bar#1{\overline{#1}}

\def\boxtext#1{\vbox{\hrule\hbox{\vrule\kern5pt
       \vbox{\kern5pt{#1}\kern5pt}\kern5pt\vrule}\hrule}}

\begin{document}
\title{\Large \bf {Hadronic interactions from effective chiral}\\
{\Large \bf Lagrangians of quarks and gluons
\footnote{To appear in the proceedings of {\em Hadron Physics 1996}, Angra dos
Reis~-~RJ,~Brazil, 15~-~20~April 1996 (World Scientific, Singapore 1996).}
}}
\author{G. Krein}
\address{Instituto de F\'{\i}sica Te\'orica, Universidade Estadual Paulista \\
\it Rua Pamplona, 145 - 01405-900 S\~ao Paulo,SP - Brazil} 
\date{}
\maketitle\abstracts{
Effective chiral Lagrangians involving constituent quarks, Goldstone bosons and
long-distance gluons are believed to describe the strong 
interactions in an intermediate energy region between the confinement scale 
and the chiral symmetry breaking scale. Baryons and mesons in such a 
description are bound states of constituent quarks. We discuss 
the combined use of the techniques of effective chiral field theory and of the 
field theoretic method known as Fock-Tani representation to derive effective 
hadron interactions. The Fock-Tani method is based on a change of 
representation by means of a unitary transformation such that the composite 
hadrons are redescribed by elementary-particle field operators. Application 
of the unitary transformation on the microscopic quark-quark interaction 
derived from a chiral effective Lagrangian leads to chiral effective 
interactions describing all possible processes involving hadrons and their 
constituents. The formalism is illustrated by deriving the one-pion-exchange 
potential between two nucleons using the quark-gluon effective chiral 
Lagrangian of Manohar and Georgi. We also present the results of a study of
the saturation properties of nuclear matter using this formalism.}

\vspace{0.5cm}


\section{Introduction}

The quark-gluon description of the interactions among hadrons and the 
properties of high temperature and/or density hadronic matter is one of 
the most central problems of contemporary particle and nuclear physics. Such 
problems are characterized by processes that involve the simultaneous presence
of hadrons and their constituents. The mathematical description of the
processes requires approximations where a drastic reduction of the 
degrees of freedom is unavoidable. In this sense, one would expect
simplifications by describing the hadrons  participating in the processes
in terms of macroscopic hadron field operators, instead of the microscopic
constituent ones. At low energies the hadron-hadron interaction can be
described by an effective chiral field theory in which the quarks and 
gluons are ``integrated out" in favor of hadrons and Goldstone
bosons~\cite{{We-PhysA},{We-NN}}. At higher energies it is very likely that 
the substructure of the hadrons will play a role and another effective field
theory involving these degrees of freedom must be introduced.

There is a widespread belief that there exists an intermediate energy region
in which it makes sense to describe the strong interactions in terms of
an effective field theory of {\em constituent quarks} subject to weak color 
forces that become strong only at large separations and keep the quarks 
confined. The $u$ and $d$ constituent quarks have a mass of $m \sim 300$ MeV,
which are believed to be the result of the spontaneous breakdown of the
$SU(2) \otimes SU(2)$ chiral symmetry. If this is so, the Goldstone bosons of 
the spontaneous symmetry breakdown (pions in the case of $u$ and $d$ quarks 
only) must be included among the degrees of freedom of the effective theory.
The lowest order terms of the Lagrangian of such an effective field theory 
were written down by Manohar and Georgi (MG)~\cite{ManoharGeorgi}. Many of the
successes of the simple nonrelativistic quark model can be understood
in this framework with a chiral symmetry breaking scale
$\Lambda_{\chi{\rm SB}} \sim 1$ GeV, which is significantly larger than
the confinement scale $\Lambda_{\rm conf} \sim 200$ MeV. 

The description of the hadron-hadron interaction in such a theory becomes
complicated because hadrons are not the basic degrees of freedom of the
theory; hadrons are composites and in general cannot be described by field 
operators of the sort used to describe elementary particles. In this talk 
we discuss a method we believe can be very useful for treating composite 
hadron interactions at the quark-gluon level. The method is known as 
Fock-Tani (FT) representation and was invented independently by 
Girardeau~\cite{girar1} and Vorob'ev and Khomkin~\cite{russ} 
to deal with atomic systems where atoms and electrons are simultaneously 
present in the system and the internal degrees of freedom of atoms cannot 
validly be neglected. The method is based on a change representation by 
introducing fictitious elementary hadrons in close correspondence to the 
real hadrons. The change of representation is implemented by  means of a 
unitary transformation such that the composite hadrons are redescribed by 
elementary-particle field operators; all field operators representing quarks, 
antiquarks, gluons and {\em hadrons} satisfy canonical commutation relations 
and therefore the traditional methods of quantum field theory can be readily 
applied. In the new representation the microscopic interquark forces change, 
they become weaker, in the sense they cannot bind the quarks into hadrons, 
and one obtains effective interactions describing all possible processes 
between hadrons and their constituents. 

Recently the original FT formalism was extended to hadronic physics for
deriving effective hadron Hamiltonians in constituent quark
models~\cite{HKSV}. Here we discuss its application in the context of the
MG effective chiral field theory. In the next section we 
discuss the main features of the formalism and the properties 
of the effective Hamiltonians. In section 3 we consider baryons in the 
MG model and derive an effective chiral Hamiltonian for baryons. 
One particularly important component present in the effective nucleon-nucleon 
interaction is the one-pion exchange interaction. We also discuss the 
saturation properties of nuclear matter in this model. In the last
section we present conclusions and discuss future perspectives.

%
\section{Fock-Tani representation: real and ideal particles}

In this section we discuss very briefly the main features of the FT 
representation. We consider a Hamiltonian where quarks and antiquarks interact
by two-body forces, and consider mesons and baryons as bound states of a 
quark-antiquark pair and three quarks respectively. In Fock space ${\cal F}$,
mesons and baryons can be written in terms of creation operators as:
\beq
M_{\alpha}^{\dag}|0\rangle \equiv \Phi_{\alpha}^{\mu \nu}
q_{\mu}^{\dag} {\bar q}_{\nu}^{\dag}|0\rangle,\hspace{1.0cm}
B_{\alpha}^{\dag}|0 \rangle  = \frac{1}{\sqrt{3!}}\,
\Psi_{\alpha}^{\mu\nu\sigma}q_{\mu}^{\dag}q_{\nu}^{\dag}
q_{\sigma}^{\dag}|0\rangle,
\label{states}
\eeq
where $\alpha$ represents the hadron quantum numbers (c.m. momentum, internal 
energy, spin and flavor), and $\Phi$ and $\Psi$ are respectively the Fock 
space meson and baryon amplitudes, where $\mu, \nu, \cdots $ represent spatial,
color, spin, and flavor quantum numbers of the quarks and antiquarks. A 
summation over repeated indices is implied. The quark and antiquark creation 
and annihilation operators obey standard anticommutation relations.
Using the quark anticommutation relations and assuming that
the meson wave-functions are orthonormalized one can show that the meson 
operators satisfy the following {\em noncanonical} commutation
relations:
\beq
[M_{\alpha}, M^{\dag}_{\beta}]=\delta_{\alpha \beta} - \Delta_{\alpha \beta},
\hspace{1.0cm} [M_{\alpha}, M_{\beta}]=0,
\label{Mcom}
\eeq
\noindent
where $\Delta_{\alpha \beta}= \Phi_{\alpha}^{*{\mu \nu }}
\Phi_{\beta}^{\mu \sigma }\bar q^{\dag}_{\sigma}\bar q_{\nu}
+ \Phi_{\alpha}^{*{\mu \nu }}
\Phi_{\beta}^{\rho \nu}q^{\dag}_{\rho}q_{\mu}$.
In addition, one can also show that meson and quark operators do not commute.
Baryon operators satisfy similar expressions~\cite{HKSV}.

One observes that the composite nature of the mesons is manifested by the term 
$\Delta_{\alpha \beta}$ in Eq.~(\ref{Mcom}), and the nonzero value for the
meson-quark commutator indicates that mesons and quarks are not independent 
degrees of freedom. The presence of these terms complicates the direct 
application of field theoretic techniques such as Wick's theorem and Feynman 
graphs for the composite hadron operators, since these techniques are set up 
for canonical field operators.  

The change to the FT representation is implemented by means of a
unitary transformation $U$, such that the {\em single} composite hadron
states are transformed into {\em single} ideal-hadron states 
$m^{\dag}_{\alpha}|0) \equiv U^{-1}M^{\dag}_{\alpha}|0\rangle$ and 
$b^{\dag}_{\alpha}|0) \equiv U^{-1}B^{\dag}_{\alpha}|0\rangle$, where the 
ideal hadron operators satisfy canonical (anti)commutation relations:
\beq
[m_{\alpha}, m^{\dag}_{\beta}]=\{b_{\alpha}, b^{\dag}_{\beta}\}=
\delta_{\alpha\beta},\hspace{1.0cm}
[m_{\alpha}, m_{\beta}]=\{b_{\alpha}, b_{\beta}\}=
[m_{\alpha}, b^{\dag}_{\beta}]=0,
\eeq
and, by definition, the $m^{\dag}$ and $m$ ($b^{\dag}$ and $b$) commute 
(anticommute) with the quark and antiquark operators. $U$ is an unitary 
operator, which can be constructed by an iterative procedure as a power series
in the $\Phi$'s and $\Psi$'s. Details on the derivation of the generator can
be found in Ref.~\cite{HKSV}; here we simply present the general form of the
effective Hamiltonian in the new representation. From the microscopic 
quark-antiquark Hamiltonian $H$, one obtains
\beq
H_{\rm FT} \equiv U^{-1}HU = H_{q} + H_{h} + H_{h q} \;,
\label{Htrans}
\eeq
where the subindices identify the operator content of each term. The quark 
Hamiltonian $H_{q}$ involves only quark and antiquark operators. In general
it has a similar structure to the one of the microscopic quark Hamiltonian 
$H$, except that it cannot produce the hadronic bound states. This feature 
leads to the same effect of curing the bound state divergencies of the Born 
series as in Weinberg's quasi-particle method~\cite{We-quasi}: the modified 
quark interaction is unable to form hadrons, they are redescribed by the 
$H_{h}$ part of the effective Hamiltonian.

$H_{hq}$ describes quark-hadron processes as hadron breakup into quarks and 
hadron-quark scattering. In models where quarks are truly confined, these
terms contribute to free-space meson-meson processes only as intermediate 
states. However, in high temperature and/or density systems hadrons and quarks
can coexist and the breakup and recombination processes can play an important 
role.

The term involving only ideal hadron operators $H_{h}$ represents effective 
meson-meson, baryon-baryon, and baryon-meson processes. We exemplify this in 
the next section for the baryon-baryon interaction using the Manohar and 
Georgi Lagrangian density. 

%
\section{Effective chiral Hamiltonian for baryon-baryon interactions}

Now we discuss the application of the FT formalism to the effective QCD field
theory of MG~\cite{ManoharGeorgi}. As discussed in the Introduction, since the
effects of dynamical chiral symmetry breaking are included in the constituent 
quark mass the interquark forces become weaker in the effective theory. This 
allows to identify the low-lying hadrons with nonrelativistic bound states of 
the constituent quarks. Of course, there remains the problem of the confining 
forces which are difficult to access in such an approach. At this stage
probably the best attitude is simply to postulate a phenomenological effective
confining interaction. We then proceed deriving from this quark-gluon theory an
effective quark-quark interaction up to a certain order in the chiral 
expansion. Once one has the microscopic quark-quark potential and 
the ``bare" hadron bound-states obtained with the phenomenological confining
interaction, one implements the FT transformation to obtain effective 
hadron-hadron interactions. Corrections to the bare hadron bound-states 
can be calculated perturbatively in the new representation.

Let us consider just the lowest order pion-quark interaction piece of the MG 
model. At tree level, this corresponds to the standard pseudovector coupling.
Given the quark-quark interaction and the baryon wave-functions $\Psi$, one 
obtains through the lowest-order FT transformation an effective 
baryon Hamiltonian $H_b$ of the general form~\cite{HKSV}:
\beq
H_b= E_{\alpha}\,b^{\dag}_{\alpha}b_{\alpha}+
\frac{1}{2} V_{bb}(\alpha\beta;\delta\gamma)\,b^{\dag}_{\alpha}b^{\dag}_{\beta}
b_{\gamma}b_{\delta},
\label{Hb}
\eeq
where $E_{\alpha}$ is the total energy (internal plus c.m.) and
$V_{bb}$ is the effective baryon-baryon interaction, which we can write as
a sum of direct and quark-exchange parts, $V_{bb}=V^{dir}_{bb}+V^{exc}_{bb}$.
The direct term $V^{dir}_{bb}$, represented in Figure 1 below, is given by:
\beq
V^{dir}_{bb}(\alpha\beta;\delta\gamma)=9\, V_{qq}(\mu \nu; \sigma \rho)
\Psi^{\ast\mu\mu_2\mu_3}_\alpha\Psi^{\ast\nu\nu_2\nu_3}_{\beta}
\Psi^{\rho\nu_2\nu_3}_{\gamma}
\Psi^{\sigma\mu_2\mu_3}_{\delta}.
\label{Vbbdir}
\eeq

\begin{figure}[h]
\vspace{-0.5cm}
\epsfxsize=4.0cm
\centerline{\epsfbox{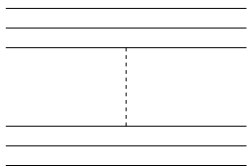}}
\vspace{-2.0cm}
\caption{Graph of $V^{dir}_{bb}$.}
\vspace{0.3cm}
\end{figure}

When the baryons have the quantum numbers of nucleons and the microscopic
quark-quark interaction is the one-pion-exchange, the explicit form of
this term is the familiar one-pion-exchange potential in momentum space:
\beq
V^{dir}_{NN}(q)= - 9\, \frac{25}{81}\, \left(\frac{g_A}{f_\pi}\right)^2\,
\tau^{a(2)}_N\tau^{a(1)}_N\, F(q^2) \frac{{\boldsigma}^{(1)}_N\bcdot{\bf q}\,
{\boldsigma}^{(2)}_N\bcdot{\bf q}}{{\bf q}^2 + m_{\pi}^2} F(q^2),
\eeq
where the $\tau^a_N, a=1, \cdots, 3 $ and ${\boldsigma}_N$ are respectively the
nucleon isospin and spin Pauli matrices, and $F(q^2)$ is the nucleon (matter) 
form factor. 

The exchange term $V^{exch}_{bb}$ involves the simultaneous exchange of pions 
between two quarks and the exchange of quarks between two baryons. For 
illustrative purposes we show explicitly only one of such terms: 
\begin{equation}
V^{exch}_{bb}(\alpha\beta;\delta\gamma)=-9 \; V_{qq}(\mu \nu; \sigma \rho)
\Psi^{\ast\mu\mu_2\mu_3}_\alpha\Psi^{\ast\nu\nu_2\nu_3}_{\beta}
\Psi^{\sigma\nu_{2}\nu_{3}}_{\gamma}
\Psi^{\rho\mu_{2}\mu_{3}}_{\delta}.
\label{Vbbexch}
\end{equation}
This is schematically represented in Figure 2. Because exchange terms involve 
quark exchange between the baryons, they are of shorter range than the one of 
$V^{dir}_{bb}$. 

\begin{figure}[h]
\centerline{
{\epsfxsize=4.0cm\epsfbox{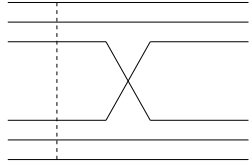}}
} 
\vspace{-2.0cm}
\caption{Graph of $V^{exch}_{bb}$.}
\vspace{0.3cm}
\end{figure}

The multidimensional integral over the quark coordinates in 
$V^{exch}_{bb}$ 
cannot in general be done explicitly. However, when the quark wave functions 
are nonrelativistic gaussians, $V^{exch}_{bb}$ can be expressed 
in terms of a single integral:
\bea
V^{exch}({\bf p},{\bf p}') &=& 
{\cal O}^{ij}(\sigma,\tau)\left(\frac{g_A}{f_\pi}\right)^2\,
\left(\frac{3b^2}{4\pi}\right)^{3/2}\,
e^{-b^2[5({\bf p}^2+{\bf p}'^2)/12-{\bf p}{\bcdot}{\bf p}'/2]}\nonumber\\
&\times & \int d^3 q \,  \frac{q^iq^j}{{\bf q}^2 + m_{\pi}^2} \,
e^{-b^2[{\bf q}^2-{\bf q}{\bcdot}({\bf p}'-{\bf p})]},
\label{V2}
\eea
where $b$ is the r.m.s. radius of the nucleon and 
${\cal O}^{ij}(\sigma,\tau)$ is given by:
\begin{eqnarray}
{\cal O}^{ij}(\sigma,\tau)&=&\frac{1}{9} \left\{\left[25+\frac{1}{3}
\tau^{a(2)}_N\tau^{a(1)}_N\left(1+18{\boldsigma}^{(1)}_N
{\bcdot}{\boldsigma}^{(2)}_N\right)\right]\delta^{ij} \right. \nonumber \\
&+&\left.
\left(1+19\tau^{a(2)}_N\tau^{a(1)}_N\right)\sigma^{i(1)}_N\sigma^{j(2)}_N
\right\}.
\end{eqnarray}

An interesting feature of  a potential derived from a quark model is that
when interated to obtain the scattering amplitude, there is no need for 
regularization and renormalization of the scattering equation, because of the 
form factors that come from the baryon wave functions. Because of this one can 
expect that the study of the baryon-baryon scattering at low energies in the 
present approach will be much easier than in the approach at the hadronic 
level, where the renormalization of the Lippman-Schwinger equation must be
performed~\cite{Bira}. 

Beyond tree-level, loops of quarks, pions and gluons introduce divergencies and
higher dimensional operators must be included to cancel the
divergencies. The new coefficients that come along with the higher dimensional
operators are fitted to experiments. Work is in progress where the four-point 
interactions and the one gluon-exchange at the tree level are summed to the 
tree-level pion exchange just discussed. The aim is to calculate phase-shifts 
of the nucleon-nucleon scattering and compare with the traditional one-boson 
exchange models.

%
\section{Nuclear matter}

Once one has an effective NN interaction, the traditional nuclear many-body
techniques are readily applicable because all field operators in the new
representation satisfy canonical (anti)commutation relations. We illustrate 
this by outlining a calculation of the equation of state of symmetric nuclear
matter in the Hartree approximation~\cite{krein}. 

If one neglects quark-exchange contributions to the effective NN interaction, 
and considers only the Hartree approximation, the relevant terms in the MG 
Lagrangian~\cite{ManoharGeorgi} are:
\beq
{\cal L} = {\cal L}_{conf} + \bar\psi\left(i \gamma^{\mu} \partial_\mu - 
m_q \right)\psi - \frac{C_s}{2f^2_\pi}(\bar\psi\psi)^2 + \frac{C_v}{2f^2_\pi}(\psi^{\dag}\psi)^2,
\label{Rel-Lag} 
\eeq
since the one-pion and one-gluon exchanges, as well as the other four-point 
interactions involving $\gamma_5$ and $\sigma^{\mu\nu}$, average to zero.
In Eq.~(\ref{Rel-Lag}), $f_\pi$ is the pion decay constant.
 
Using the Fermi-Breit approximation for the effective four-quark interactions,
one obtains an effective NN interaction which gives an energy density 
$E/A-M_N={\cal E}/\rho - M_N$ for symmetric nuclear matter of the 
form~\cite{krein}:
\begin{eqnarray}
\frac{E}{A}-M_N &=& 3(m_q-m^*_q)+\frac{3}{5}\,\frac{k_F^2}{6m^*} + 
\frac{f^2_\pi}{18C_s\rho}\,(3m_q-3m^*_q)^2+\frac{9C_v}{2f^2_\pi}\,\rho 
\nonumber\\
&+& \frac{3}{2m_qb^2}\,\left(\frac{b^{*2}}{b^2}+\frac{m_q}{m^*_q}
\frac{b^2}{b^{*2}}-2\right),
\end{eqnarray}
where $M_N$ is the nucleon mass (calculated within the model), 
$\rho=2k^2_F/3\pi^2$ is the nuclear density and $k_F$ is the nuclear Fermi 
momentum, and the effective quark mass $m^*_q$ satisfies the self-consistent 
equation:
\beq
3m^*_q=3m_q-\frac{9C_s}{f^2_\pi}\left[\left(1-\frac{1}{2m^{*2}b^{*2}}\right)
- \frac{3}{5}\frac{k^2_F}{2(3m^*_q)^2}\right]\rho.
\eeq
The confining interaction is taken to be an harmonic oscillator. Note that the 
in-medium size of the nucleon $b^*$ is allowed to be different from $b$, its
value is determined self-consistently with $m^*_q$ by requiring that $E/A$ be 
a minimum also with respect to variations on $b^*$. 

Nuclear matter can be saturated with $E/A-M_N=-15.64$ MeV at 
$k_F~=~1.364$~fm$^{-1}$ using the following set of parameters: 
$9C_s/2f^2_\pi~=~8.09$,  $9C_v/2f^2_\pi~=~0.56$,  $b~=~0.6fm$, 
$m_q~=~350$~MeV. At the saturation point, we find $m^*_q/m_q~=~0.81$ and 
$b^*/b~=~1.05$. The compressibility is found to be~$150$~MeV.

%
\section{Conclusions and future perspectives}

The combined use of the techniques of effective chiral field theory and the 
Fock-Tani representation seems to provide great opportunities to study the 
role of quarks and gluons in hadronic interactions. The fact that in the
Fock-Tani representation all operators satisfy canonical (anti)commutation 
relations allows the direct use of the known field theoretic techniques such 
as Feynman diagrams and Green's functions which have proven to be very useful 
in the study of processes involving elementary particles. Particularly 
interesting applications of these techniques are the study of short-range 
hadron-hadron interactions and the problem of hadron properties in a hot 
and/or dense medium. 


\end{document}